\documentclass[sigconf]{acmart}
\AtBeginDocument{%
  }

\usepackage{algorithm} 
\usepackage{algpseudocode} 
\usepackage{xcolor}
\usepackage{tikz} 
\usetikzlibrary{patterns}
\usetikzlibrary{calc}
\usepackage{tikz-3dplot}
\usetikzlibrary{arrows.meta}
\usepackage{wrapfig} 

\usepackage{morefloats}

\usepackage{seqsplit}

\usepackage{xspace}
\newcommand{\sysname}{\textsc{Pierce}\xspace}

\copyrightyear{2026}
\acmYear{2026}
\setcopyright{cc}
\setcctype{by}
\acmConference[SIGSPATIAL '26]{The 34th ACM International Conference on Advances in Geographic Information Systems}{November 03--06, 2026}{Riverside, CA, USA}
\acmBooktitle{The 34th ACM International Conference on Advances in Geographic Information Systems (SIGSPATIAL '26), November 03--06, 2026, Riverside, CA, USA}
\acmDOI{10.1145/3841645.3844195}
\acmISBN{979-8-4007-2950-8/2026/11}

\begin{document}

\title{Pierce: GPU Ray Tracing for Spatial Joins over Complex 3D Data}

\author{Anton Hackl}
\orcid{0009-0000-4996-8934}
\affiliation{%
  \institution{Hasso Plattner Institute, University of Potsdam}
  \city{Potsdam}
  \country{Germany}
}
\email{anton.hackl@student.hpi.de}

\author{Eleni Tzirita Zacharatou}
\orcid{0000-0001-8873-5455}
\affiliation{%
  \institution{Hasso Plattner Institute, University of Potsdam}
  \city{Potsdam}
  \country{Germany}
}
\email{eleni.tziritazacharatou@hpi.de}

\renewcommand{\shortauthors}{A. Hackl and E. Tzirita Zacharatou}

\begin{abstract}
Many emerging applications, from computational biology to digital twins and urban planning, rely heavily on three-dimensional spatial joins over polyhedral meshes. 
These joins comprise computationally intensive triangle--triangle intersection tests that pairwise compare the faces of polyhedral meshes. 
Since each mesh may contain thousands of faces, the resulting cost challenges the responsiveness of spatial data management techniques. Existing techniques follow the filter-and-refine paradigm, accelerating either the filtering step through indexing or the refinement step through progressive mesh compression combined with GPU parallelization of triangle--triangle tests. 
However, the former neglects the high cost of intra-geometry refinements, whereas the latter lowers this cost but still relies on the same pairwise triangle--triangle tests.
In this paper, we introduce \sysname, an approach that reformulates three-dimensional spatial joins over complex polyhedral meshes as ray-tracing operations and leverages the hardware ray-tracing units (RT cores) of modern GPUs to accelerate query execution.
Our approach casts rays along the edges of one mesh against a spatial hierarchy built over the other, performing ray--node tests at the internal levels to prune distant geometries and ray--triangle tests at the leaves to identify intersecting meshes, both of which RT cores accelerate in hardware.
We evaluated \sysname on real and synthetic data against multiple baselines,
demonstrating more than two orders of magnitude speedup on digital pathology data compared to the state-of-the-art approach.
\end{abstract}



\maketitle



\section{Introduction}
Many scientific and engineering applications represent physical structures as collections of complex three-dimensional objects, each modeled as a polyhedral mesh of triangular faces~\cite{schneider2002geometric}.
In computational biology, serial-section electron microscopy reconstructs brain tissue into large collections of such meshes, including cell nuclei, blood vessels, and axon bundles~\cite{microns}. 
Understanding their spatial relationships helps reveal, for example, which nuclei lie within the tissue irrigated by a given set of vessels~\cite{tdbase-simulator, 3dpro, stitch}.
In urban planning, 3D city models, such as CityGML, represent buildings, terrain, and infrastructure as meshes whose spatial overlaps drive flood simulation and sunlight analysis~\cite{groger2012citygml, park2021flood, chaturvedi2017solar}. 
In manufacturing, CAD assemblies comprise thousands of mechanical parts modeled as meshes, whose interpenetrations can indicate interference or tolerance violations~\cite{jimenez2001collision, lin1998collision}.
Underlying all these domains is a common operation, the 3D spatial join, which identifies pairs of spatially related objects across two datasets.
Analyses such as those mentioned above are often exploratory and interactive, necessitating low query latency to enable domain experts to investigate the data more effectively~\cite{liu2014latency}.

\noindent \textbf{Challenges.} 
Existing techniques for evaluating 3D spatial joins use a two-step filter-and-refine strategy~\cite{guting1994spatial}. 
The filter step builds a spatial index over coarse object approximations, such as minimum bounding boxes, to prune object pairs that do not satisfy the join predicate.
The refinement step then performs exact geometric tests on the original meshes to discard false positives. 
While filtering over coarse approximations is generally cheap, refinement is computationally intensive, as it involves testing whether any face of one mesh intersects a face of the other.
A single real-world mesh can have tens of thousands of faces, leading to millions of tests for even one candidate pair. 
With many such pairs in a dataset, the refinement cost compounds rapidly, dominating query time and making fast response times difficult to achieve.

\noindent \textbf{Limitations of Existing Work.} 
Existing work on spatial joins addresses this bottleneck from two directions.
The first direction strengthens the filtering step by using spatial indexes~\cite{tsitsigkos2019parallel, nobari2013touch} to efficiently discard candidate pairs.
However, these index-based methods treat each surviving pair as a black box and lack mechanisms to reduce the refinement cost, which is the dominant cost for complex 3D meshes~\cite{3dpro}.
The second direction reframes the refinement step. 
The state-of-the-art approach, TDBase~\cite{tdbase}, and its predecessor, 3DPro~\cite{3dpro}, extend the filter-and-refine paradigm into \emph{filter-and-progressive-refine}: each polyhedron is represented at multiple levels of detail (LODs), and every candidate pair is refined from coarse to fine LOD, terminating as soon as some LOD resolves the predicate. 
This trades a single full-resolution refinement for several cheaper rounds, betting that most pairs resolve early. 
However, each round still executes several pairwise triangle--triangle tests on general-purpose cores.

\noindent \textbf{Contributions.} 
This paper introduces \textsc{Pierce}, an approach that reformulates 3D spatial joins over polyhedral meshes as ray-tracing operations, mapping the join's core computation onto primitives that modern GPUs accelerate in dedicated hardware ray-tracing units (RT cores). 
Our approach builds on a key geometric insight: when the surfaces of two closed polyhedral meshes cross, at least one edge of one mesh must pierce a face of the other~\cite{interference_detection}.
This insight allows us to reduce triangle--triangle comparisons to ray-tracing operations.
Specifically, \textsc{Pierce} casts a ray along each edge of one dataset and traces it against a spatial hierarchy built over the triangular faces of the other, identifying which faces each ray pierces.
Ray tracing against the spatial hierarchy prunes regions a ray cannot reach at internal levels (filtering) and tests primitives at leaves (refinement), all using dedicated ray-tracing hardware. 
The key distinction from prior work lies in the primitive used: while conventional refinement compares \emph{pairs of triangles}, \textsc{Pierce} tests a \emph{ray against a triangle}, which is precisely the operation RT cores are optimized for.

In summary, we make the following contributions:
\begin{itemize}
  \item We show that 3D spatial joins can be reformulated as ray-tracing operations, which is precisely the workload that modern GPUs accelerate in dedicated hardware (RT cores). Building on this insight, we introduce 3D spatial join operators accelerated by RT cores, reinterpreting the traditional filter-and-refine paradigm (Section~\ref{sec:approach}). 

\item
\begingroup
\setlength{\emergencystretch}{1em}
We design an edge-ray strategy that detects all surface crossings between
two datasets through two complementary mechanisms:
\emph{bidirectional tracing} casts rays from both datasets in turn, so a
crossing is found regardless of which mesh contributes the piercing edge,
while \emph{multi-hit traversal} records every crossing along a ray, so
that crossings with even the most distant objects are found
(Section~\ref{sec:approach}).
\par
\endgroup

  \item We develop algorithms for three join predicates: overlap, containment, and
  intersection (Sections~\ref{sec:overlap}--\ref{sec:intersection}). Furthermore, we introduce a selectivity estimator that predicts the size of the join result in advance. This allows the result structure to be allocated at the right size on the GPU, maintaining query performance without wasting GPU memory (Section~\ref{sec:result-collection}).
  
  \item We evaluate \textsc{Pierce} on real and synthetic data against multiple
  baselines, including the state-of-the-art GPU-accelerated TDBase~\cite{tdbase}, and show that it
  achieves the lowest query time across all workloads, with over two orders of magnitude speedup over TDBase on digital pathology data (Section~\ref{sec:experimental-evaluation}).

\end{itemize}

\section{Background: Ray Tracing}

\begin{figure}[t]
    \centering

    \begin{minipage}[t]{0.54\linewidth}
        \centering
        \includegraphics[width=\linewidth, angle=0,
    origin=c]{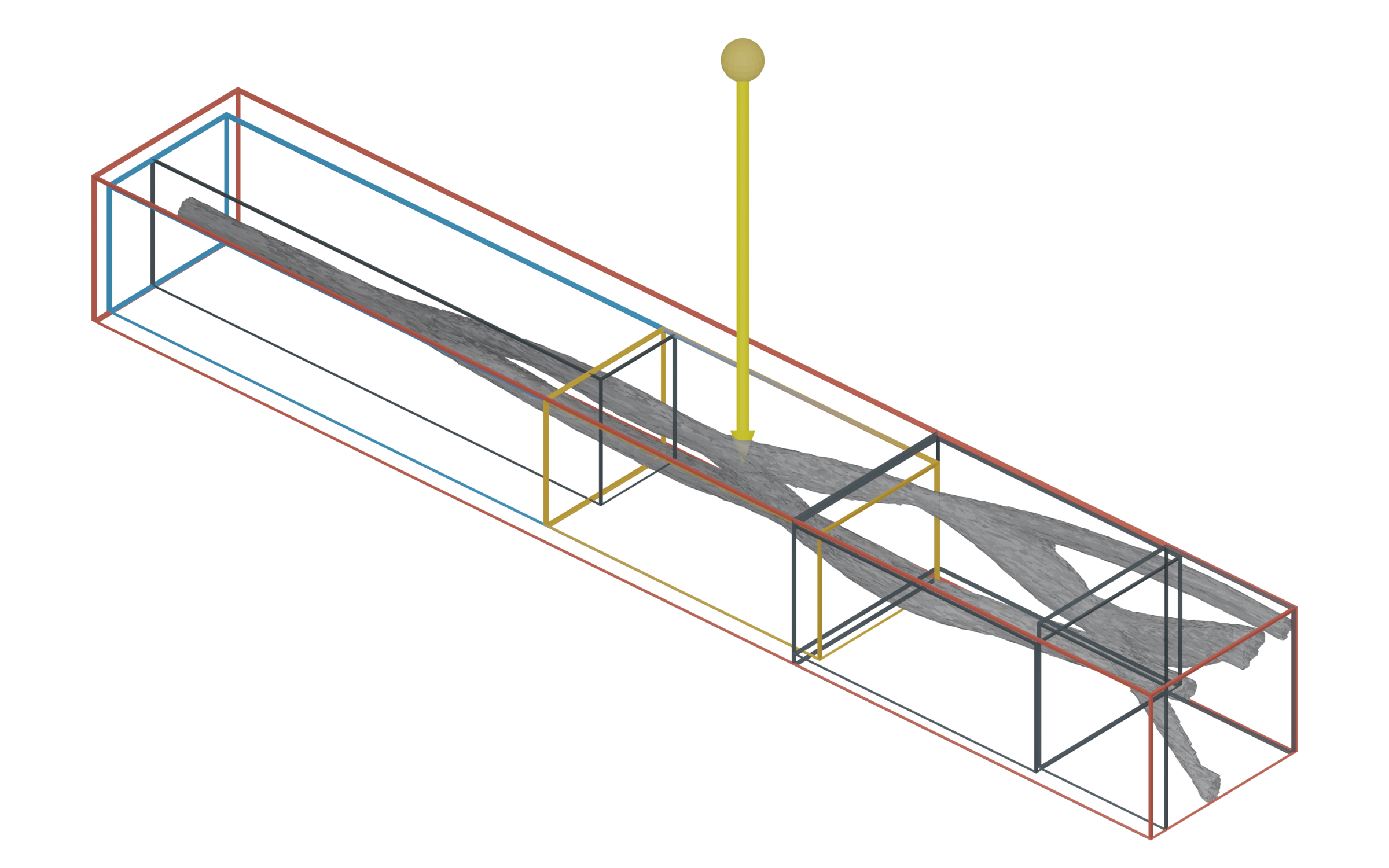}
    \end{minipage}\hfill
    \begin{minipage}[t]{0.40\linewidth}
        \centering
        \includegraphics[width=\linewidth]{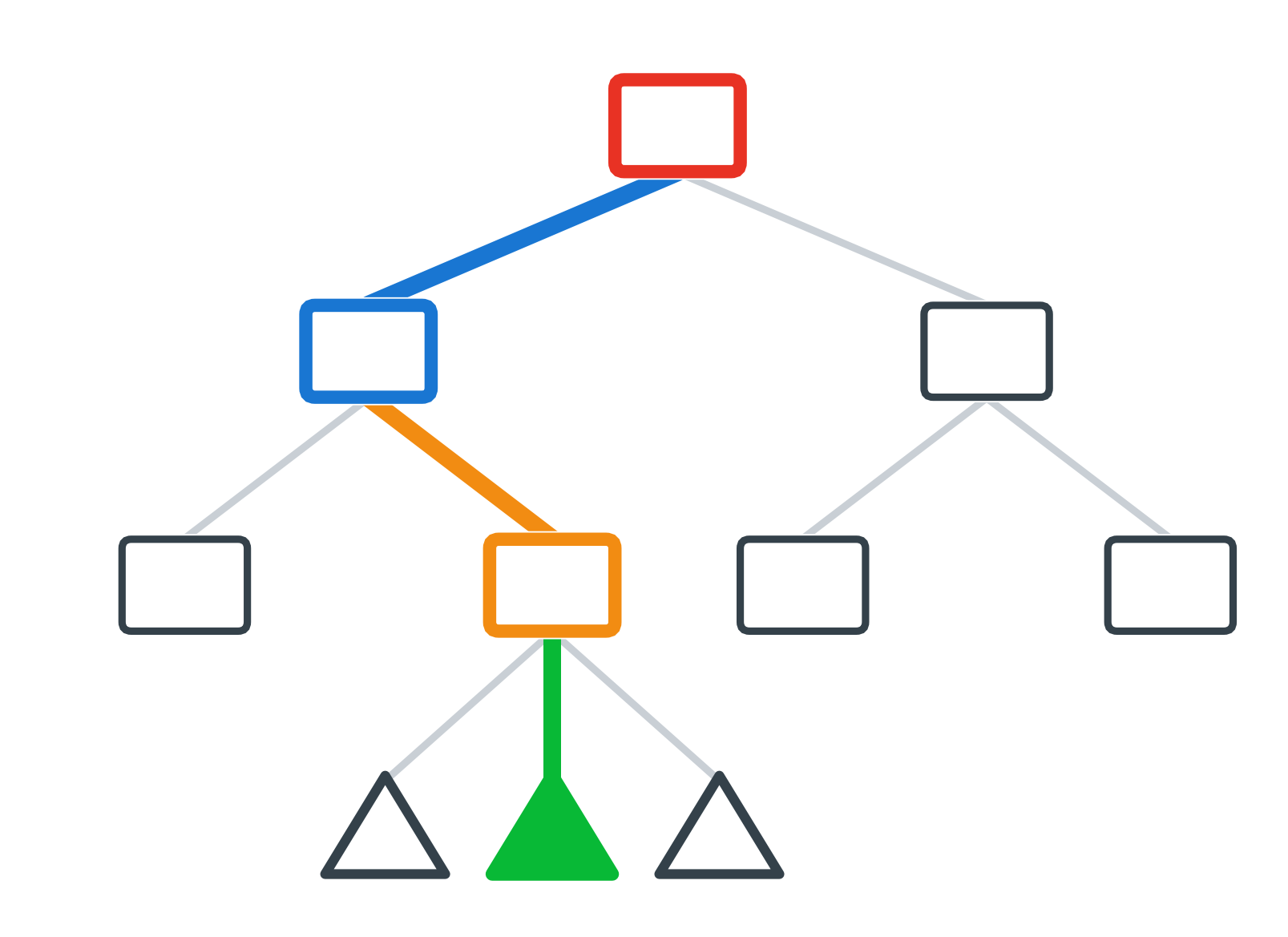}
    \end{minipage}

    \caption{Ray--triangle intersection with BVH traversal. A query ray is traced against the scene geometry (left), while the corresponding BVH traversal prunes non-intersected regions and descends to the intersected triangle set (right).}
    \Description{Two panels illustrate BVH pruning for a vessel-like triangle mesh. The left panel shows a downward query ray intersecting the mesh within nested bounding boxes; only the boxes along its path need further examination. The right panel shows the corresponding hierarchy, where a highlighted path descends from the root through selected internal nodes to one leaf triangle, while sibling branches are skipped.}
    \label{fig:bvh}
\end{figure}

Ray tracing is a rendering technique that simulates the physical behavior of light by casting rays from a viewpoint into a scene and determining the surfaces they intersect. Each intersection reveals what is visible along the ray and may trigger secondary rays for effects such as reflection or shadow. The central computational challenge is therefore to find which surfaces a given ray intersects.

To tackle this efficiently, 3D objects are represented as triangle meshes, and a Bounding Volume Hierarchy (BVH) is constructed over the triangles in a scene. The BVH is a tree in which each internal node stores an axis-aligned bounding box enclosing all triangles in its subtree and each leaf holds a small number of triangles. It is closely related to the R-tree used in spatial databases~\cite{guttman1984rtrees}.

Ray tracing is accomplished by traversing the BVH, as illustrated in Figure~\ref{fig:bvh}. 
 The left side shows a triangulated vessel model, while the right displays the corresponding BVH, with colored bounding boxes matching the tree nodes. 
To trace a ray against the scene, modern GPUs utilize dedicated \emph{RT cores} that traverse the BVH top-down, pruning branches whose bounding boxes the ray misses and performing ray--triangle intersection tests only at the leaves. 
In our example, the yellow ray is traced by following the highlighted path from the root (red) through the blue internal node to the orange leaf, where each enclosed triangle is tested to identify the intersecting green one.

The NVIDIA OptiX API~\cite{nvidia2024optix} exposes a programmable ray-tracing pipeline built on top of hardware BVH traversal, providing \sysname with both an acceleration structure and a set of programmable traversal callbacks. In our implementation, each dataset is represented as a set of triangle primitives, over which OptiX builds a single geometry acceleration structure (GAS), the BVH that the RT cores traverse. 
Traversal is then customized by user-defined programs invoked at fixed points in the pipeline. 
A \textit{ray-generation} program launches the rays, a \textit{closest-hit} program runs at the nearest intersection along a ray, and an \textit{any-hit} program runs at every intersection. 
\sysname specializes these programs to map each 3D spatial join predicate onto a ray-tracing query (see Section~\ref{sec:exp:optix}).


\section{Ray-Tracing-Based 3D Spatial Join}
\label{sec:approach}

Evaluating 3D spatial join predicates between polyhedral objects involves testing the geometric relationships between their mesh primitives. 
Equivalently, the join can be framed as casting rays from one mesh and observing where they pierce the surfaces of the other. 
This task maps directly onto an operation that modern GPUs can efficiently handle using dedicated ray-tracing hardware: ray--triangle intersection over a spatial hierarchy (i.e., a BVH) of triangles. 
Instead of the conventional two-stage filter-and-refine pipeline, the join can be expressed as a series of ray-tracing queries, where each hardware-accelerated BVH traversal handles both filtering and refinement.

This section presents \sysname, a 3D join framework built on this idea.
Section~\ref{sec:core-approach} describes the core approach.
We then develop three join operators on top of this approach: the \emph{overlap join} (Section~\ref{sec:overlap}) detects surface crossings; the \emph{containment join} (Section~\ref{sec:containment}) returns pairs where one object fully encloses the other, with no surface crossings; and the \emph{intersection join} (Section~\ref{sec:intersection}) returns pairs that satisfy either predicate.
Section~\ref{sec:result-collection} describes how the resulting object pairs are materialized, while Section~\ref{sec:discussion} discusses extensions.

\begin{figure}[t]
  \centering
  \includegraphics[width=\linewidth]{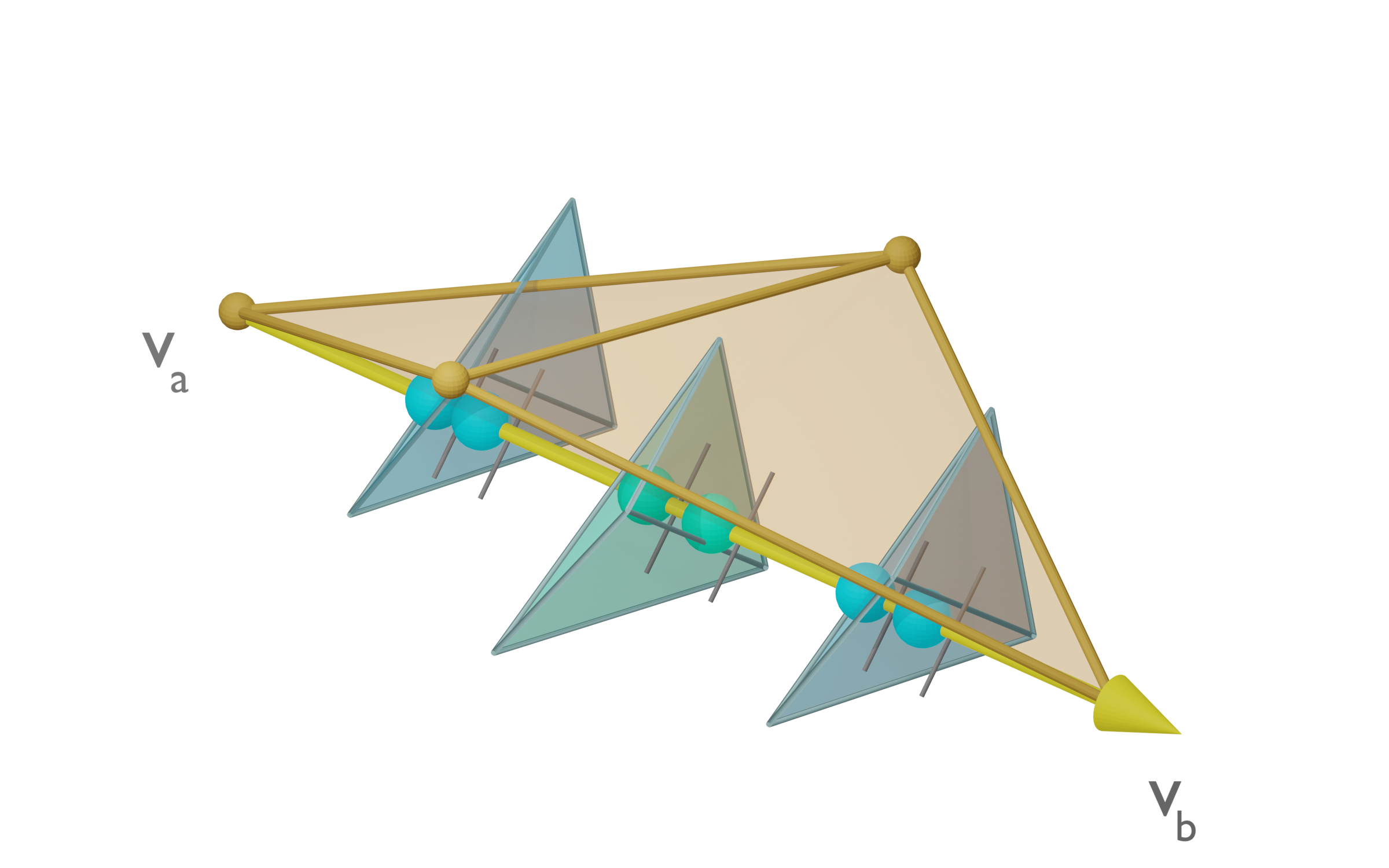}
  \caption{
    Visualization of the ray-tracing-based join primitive.
    The orange mesh represents a source object, and the cyan meshes represent target objects.
    A bounded ray is cast along the source edge from $v_a$ to $v_b$, shown in yellow, and intersects multiple target objects along its length.
    Colored spheres mark the hit locations on the edge, while the small dark tick marks indicate how the ray interval is advanced to the next representable floating-point distance after each hit to continue searching for intersections along the same edge.
  }
  \Description{A single bounded ray follows an edge of the source triangle and crosses three separate target triangles. Each crossing is recorded, and then the ray origin advances by the smallest representable distance beyond the hit, allowing the same edge to find all subsequent intersections rather than stopping at the first.}

  \label{fig:approach}
\end{figure}

\subsection{Core Approach}
\label{sec:core-approach}

The design of \sysname builds on two key observations:

\begin{enumerate}
    \item \textit{When two closed polyhedral surfaces cross, at least one edge of one mesh must pierce a face of the other} \cite{interference_detection}. Detecting every such edge--face crossing between two datasets is therefore sufficient to detect all overlapping object pairs.
    \item \textit{Tracing a ray cast along an edge against a BVH over the other dataset's triangles simultaneously prunes distant objects and performs exact ray--triangle intersection tests at the leaves.} As a result, a single hardware-accelerated BVH traversal effectively replaces both steps of the traditional filter-and-refine approach.
\end{enumerate}

Given two datasets $D_{source}$ and $D_{target}$ of 3D polyhedral meshes, \sysname organizes all triangles of each dataset into a single BVH and casts rays along the unique edges of $D_{source}$ against $D_{target}$'s BVH.
Each ray originates at one endpoint of the edge, points toward the other endpoint, and the ray length is equal to the edge length.
A hit within this bounded range confirms that the edge pierces a triangle of a target object (an edge--face crossing), and thus the surfaces of the two objects cross. 
Since a single edge may be shared by several source objects, the hit yields a join pair for each of them.
The following steps describe the core tracing primitive in \sysname.

\noindent \textbf{Step I: Edge ray construction.}
For each unique edge $(v_a, v_b)$ in the source dataset, \sysname constructs a ray with origin $v_a$, direction $\mathbf{d} = (v_b - v_a)/\|v_b - v_a\|$, and maximum length $t_{\max} = \|v_b - v_a\|$, as shown in Figure~\ref{fig:approach}. 
A natural alternative would be to cast rays from the center of each triangular face along its surface normal, probing whether the ray immediately exits into another object's interior. 
Edge rays have two advantages over this approach.
First, edge rays directly test the geometric event that defines surface crossing: an edge of one mesh piercing a face of another. Normal-direction probes detect crossings only indirectly, by checking whether a face's outward direction leads into another object, and can miss crossings entirely when concave or interlocking surfaces cause normals to point away from the intersection region.
Second, each edge ray has a natural, tight length bound: the ray is exactly as long as the edge, so a hit is recorded only if the intersection lies on the edge segment itself, eliminating false positives from objects further along the ray.

Interior edges of a closed manifold are shared by exactly two triangles, so a naive per-triangle scheme would unnecessarily trace each edge twice. \sysname therefore deduplicates edges during preprocessing on a per-object basis, merging duplicate edges that share the same endpoints within the same source object and annotating each unique edge with its source object identifier.

\noindent \textbf{Step II: Intersection detection.}
In this step, \sysname traces each edge ray against the target BVH using a single ray-tracing query. 
If a hit occurs within the range $[0, t_{\max}]$, it indicates that the edge pierces a triangle of the target mesh (an edge--face crossing). 
The triangle that was hit is resolved to its corresponding target object $o_t$ through a triangle-to-object mapping.
A joining pair $(o_s, o_t)$ is then recorded for the source object $o_s$ associated with the edge.
As discussed in Section~\ref{sec:overlap}, this approach can be adapted to handle multiple target objects that may lie along the same edge.

The following sections extend the core approach with multi-hit traversal to handle multiple crossings along an edge, bidirectional tracing to guarantee completeness, and parity tests for containment detection, combining these ingredients into three join predicates.

\subsection{Overlap Join}
\label{sec:overlap}

Given two sets of 3D polyhedra $D_1$ and $D_2$, the overlap join returns all pairs $(o_i, o_j) \in D_1 \times D_2$ such that the surfaces of $o_i$ and $o_j$ cross, i.e., at least one edge of one mesh pierces a face of the other. 
The overlap predicate detects only surface crossings.
It excludes the case where one object is fully contained within the other without any surface crossing, as well as purely tangential contact, where two surfaces touch along a face, edge, or vertex but neither interior enters the other (see Section~\ref{sec:discussion}).

\sysname evaluates the overlap predicate by applying the core primitive of Section~\ref{sec:core-approach}, casting edge rays from one dataset against the other's BVH.
Algorithm~\ref{alg:overlap} presents the base procedure.
For each unique edge in the source dataset, the algorithm finds all surface crossings along the edge by re-invoking the ray-tracing query with an advancing $t_{\min}$ after every hit (lines~5--15). 
Each hit triangle is resolved to its owning target object via the triangle-to-object mapping (line~10), and a crossing pair is recorded for every source object incident to the edge (lines~11--13). 
The outer loop over unique edges is processed in parallel on the GPU.

\begin{algorithm}[t]
\caption{Edge Ray Tracing (one direction: $D_s \to D_t$)}
\label{alg:overlap}
\begin{algorithmic}[1]
\Require Unique source edges $E_s$ with per-edge source-object lists $\texttt{srcObjs}$
\Require Target BVH over $D_t$, mapping $\texttt{triToObj}_t$: triangle $\to$ object
\Ensure All surface-crossing pairs $(o_s, o_t)$ detected in this direction
\ForAll{unique edges $(v_a, v_b) \in E_s$ \textbf{in parallel}}
    \State $\mathbf{d} \gets v_b - v_a$
    \State $t_{\max} \gets \|\mathbf{d}\|$; \quad $\mathbf{d} \gets \mathbf{d} / t_{\max}$
    \State $t_{\min} \gets 0$
    \Repeat \Comment{multi-hit loop}
        \State hit $\gets$ \Call{TraceRay}{$v_a$, $\mathbf{d}$, $t_{\min}$, $t_{\max}$, BVH$_t$}
        \If{hit = $\emptyset$}
            \State \textbf{break} \Comment{no more crossings along this edge}
        \EndIf
        \State $o_t \gets \texttt{triToObj}_t[\text{hit.triangleIdx}]$ \Comment{target object}
        \ForAll{$o_s \in \texttt{srcObjs}(v_a, v_b)$} \Comment{source objects sharing this edge}
            \State \Call{RecordPair}{$o_s$, $o_t$} \Comment{insert into deduplicated result set}
        \EndFor
        \State $t_{\min} \gets$ \Call{NextAfter}{$\text{hit.distance}$, $t_{\max}$} \Comment{advance to next representable distance}
    \Until{no further hits}
\EndFor
\end{algorithmic}
\end{algorithm}

\noindent \textbf{Multi-hit traversal.}
The advancing-$t_{\min}$ loop (lines~5--15) is necessary because each BVH contains triangles from \emph{all} objects in the dataset.
Therefore, a single edge may intersect the surfaces of multiple target objects along its length. 
A single closest-hit query would return only the nearest intersection, missing crossings with more distant objects. The iterative loop instead reports successive hits along the edge segment: after each hit, the ray origin is advanced past the intersection point and another ray-tracing call is performed to find the next intersection, until no further intersections are found within the edge bounds. Iterating over hits is essential for correctness whenever the target dataset contains densely packed objects whose surfaces lie along the same edge.

\noindent \textbf{Bidirectional tracing.}
Algorithm~\ref{alg:overlap} traces rays from $D_s$'s edges against $D_t$'s BVH, and therefore detects only crossings where an edge of $D_s$ pierces a face of $D_t$.
The symmetric case (an edge of $D_t$ piercing a face of $D_s$) requires tracing in the opposite direction.
To guarantee completeness, the overlap join runs Algorithm~\ref{alg:overlap} twice with source and target swapped: once with $D_s = D_1$ and $D_t = D_2$, and once with $D_s = D_2$ and $D_t = D_1$.
Together, the two passes detect all edge--face crossings regardless of which mesh contributes the piercing edge and which contributes the pierced face.
Both passes write into the same result set.

\subsection{Containment Join}
\label{sec:containment}

Given two sets of 3D polyhedra $D_1$ and $D_2$, the containment join returns all pairs $(o_i, o_j) \in D_1 \times D_2$ such that $o_i$ fully contains $o_j$, i.e., every point of $o_j$ lies inside $o_i$ with no part of $o_j$'s surface crossing $o_i$'s surface.
An object pair satisfies containment exactly when there is no surface crossing between the two objects and one object lies inside the other.
\sysname therefore extends the overlap join with one additional step: after identifying all surface-crossing pairs via bidirectional edge ray tracing as in Section~\ref{sec:overlap}, it determines for each non-crossing candidate whether one object lies inside the other using a point-in-mesh parity test~\cite{schneider2002geometric}.
Algorithm~\ref{alg:containment} presents the complete procedure.

\begin{algorithm}[t]
\caption{Containment Join}
\label{alg:containment}
\begin{algorithmic}[1]
\Require Datasets $D_1$, $D_2$ with per-dataset BVHs
\Ensure All pairs $(o_i, o_j)$ where $o_i$ fully contains $o_j$
\Statex
\Statex \textbf{Surface crossing detection}
\State $H_{\text{cross}} \gets \emptyset$ \Comment{surface-crossing set}
\State Run bidirectional edge ray tracing (Alg.~\ref{alg:overlap}) between $D_1$ and $D_2$
\State Store all crossing pairs in $H_{\text{cross}}$
\Statex
\Statex \textbf{Point-in-mesh parity test}
\State $H_{\text{contain}} \gets \emptyset$ \Comment{containment results}
\ForAll{objects $o_j \in D_2$ \textbf{in parallel}}
    \State $v \gets$ representative vertex of $o_j$
    \State $\text{parity}[\ ] \gets 0$ \Comment{per-$D_1$-object parity state}
    \State \Call{TraceRayAnyHit}{$v$, $+\hat{z}$, $0$, $\infty$, BVH$_1$}
    \Statex \qquad \Comment{\textbf{Any-hit shader for hit triangle $t$:}}
    \Statex \qquad $o_i \gets \texttt{triToObj}_1[t]$
    \Statex \qquad $\text{parity}[o_i] \gets \text{parity}[o_i] \oplus 1$ \Comment{toggle parity}
    \Statex \qquad \Call{IgnoreIntersection}{} \Comment{continue traversal}
    \ForAll{$o_i$ with $\text{parity}[o_i]$ odd}
        \If{$(o_i, o_j) \notin H_{\text{cross}}$} \Comment{no surface crossing}
            \State Insert $(o_i, o_j)$ into $H_{\text{contain}}$
        \EndIf
    \EndFor
\EndFor
\State \Return $H_{\text{contain}}$
\end{algorithmic}
\end{algorithm}

\noindent \textbf{Surface crossing detection.}
\sysname first identifies all pairs that have at least one edge--face crossing, reusing the bidirectional edge ray procedure of the overlap join (Section~\ref{sec:overlap}), and stores them in a set $H_{\text{cross}}$. These pairs cannot satisfy full containment.

\noindent \textbf{Point-in-mesh parity test.}
For each object $o_j \in D_2$, \sysname casts one fixed-direction ray from a representative vertex against $D_1$'s BVH to determine which objects in $D_1$ contain that vertex. We use $+\hat{z}$, though any direction that does not graze a face works for parity counting. The any-hit shader processes all surface crossings in one traversal, avoiding a multi-hit loop that would relaunch rays: at each hit, it resolves the owning object $o_i$, toggles its per-object parity bit, and ignores the intersection to continue tracing. After traversal, odd parity indicates that the vertex lies inside $o_i$. If $(o_i, o_j)$ is absent from $H_{\text{cross}}$, then $o_i$ fully contains $o_j$ and the pair is added to $H_{\text{contain}}$.

This test is both correct and efficient. For two closed surfaces that do not cross, $o_j$ lies inside $o_i$ if and only if a single representative vertex of $o_j$ does. An empty $H_{\text{cross}}$ entry certifies the non-crossing condition and an odd parity count indicates that the vertex is interior, so together they indicate containment, while testing one representative vertex per $D_2$ object suffices.

The \texttt{within} predicate is the symmetric inverse: $o_j$ is within $o_i$ iff $o_i$ contains $o_j$. \sysname returns the same result set with the pair roles swapped, at no additional cost.

\subsection{Intersection Join}
\label{sec:intersection}

Given two sets of 3D polyhedra $D_1$ and $D_2$, the intersection join returns all pairs $(o_i, o_j) \in D_1 \times D_2$ whose surfaces cross or where one object fully encloses the other.
The intersection join therefore combines both primitives developed in Sections~\ref{sec:overlap} and~\ref{sec:containment}.

\sysname evaluates the intersection join by running the bidirectional edge ray tracing and the point-in-mesh parity test as two independent steps.
The edge ray tracing detects all surface-crossing pairs as in the overlap join (Section~\ref{sec:overlap}).
The parity test determines, for each object, whether it lies inside any object in the other dataset, as in the containment join (Section~\ref{sec:containment}).
However, unlike containment, the two steps are independent: the parity test results are not filtered against the surface-crossing results, since an object can simultaneously surface-cross one target object and be fully enclosed by another.
Both steps write to a shared result set, and their combined output is the intersection result.

\subsection{Result Materialization}
\label{sec:result-collection}

The join algorithms of Sections~\ref{sec:overlap}--\ref{sec:intersection} generate object
pairs $(o_i, o_j)$ that must be materialized into a deduplicated result set. This step is
challenging for two reasons. 
First, the edge-ray mechanism may report the same pair $(o_i, o_j)$ multiple times for three distinct reasons. 
When the surfaces of $o_i$ and $o_j$ intersect, they typically meet along a curve rather than
at a single point, so several edges of $o_i$ pierce faces of $o_j$, each producing a separate
hit.
The symmetric reverse-direction pass may then rediscover the pair through edges of $o_j$ piercing faces of $o_i$. Finally, a single edge that enters and exits the same target object can report the pair more than once within one ray's multi-hit loop. 
Second, \sysname must determine the result structure's size before launching the tracing process, as the GPU cannot resize device allocations during execution.
If the allocation is undersized, the kernel has to abort and be relaunched with a larger allocation, which hurts performance.
However, GPU memory is limited, so oversizing the structure as a safety margin wastes a scarce resource. 
\sysname must therefore carefully choose the size of the result structure in advance to prevent kernel relaunches while avoiding memory waste.
We propose two materialization strategies to address these challenges.

\noindent\textbf{Counting two-pass strategy.} This strategy performs two passes per tracing
direction. In the first pass, each thread follows its edge ray and counts the records it
would write, without writing them. A parallel prefix sum over the per-edge counts then
yields the write offsets into a flat output buffer. In the second pass, each thread retraces
its edge ray and emits the corresponding $(o_i, o_j)$ records starting at its computed
offset. Sorting the buffer and extracting unique entries finally yields the deduplicated
result set. This strategy requires no estimate of the result size, but it traces each ray
twice and adds a sort-based deduplication step.

\noindent\textbf{Estimation-based single-pass strategy.} This strategy eliminates the need for a second pass. During a single tracing pass per direction, each thread inserts the discovered
pairs $(o_i, o_j)$ directly into a pre-allocated GPU hashmap that resolves duplicates on
insertion, ensuring uniqueness. However, pre-allocating the hashmap requires fixing its capacity before execution, so this strategy requires an accurate estimate of the result size. 
An undersized hashmap increases the length of collision chains and degrades performance (and may even force a kernel relaunch if it overflows), whereas an oversized one wastes GPU memory.

To determine the hashmap's size, \sysname estimates the number of intersecting pairs during preprocessing.
To achieve this, it overlays both datasets with a regular grid and computes, for each cell, the expected number of intersecting pairs $(o_i, o_j)$ with $o_i \in D_1$ and $o_j \in D_2$.
To compute this estimate, we maintain the following statistics within each cell, for each dataset $k \in \{1, 2\}$: the number $N_k$ of objects from $D_k$ whose axis-aligned bounding box (AABB) touches the cell, the mean AABB side length $S_k$ of those objects (averaged over width, height, and depth), and their mean mesh-to-AABB volume ratio $\mathrm{VolRatio}_k$.

To estimate the probability that two objects in the same cell intersect, we treat each object
as a cube whose side length equals its mean AABB extent. 
Two such cubes overlap iff the displacement between their centers lies within a cube of side $S_1 + S_2$, known as the Minkowski sum~\cite{lozano1983spatial} of the two cubes. 
We call the volume of this cube the \emph{interaction volume}:
\[
    V_{\text{int}} = (S_1 + S_2 + \varepsilon)^3,
\]
where $\varepsilon$ (default $10^{-3}$) is a small additive bias that keeps the interaction volume positive for degenerate or near-zero-extent objects. 

Assuming the two object centers are independently and uniformly distributed within the cell, we approximate the probability that the cubes overlap by the ratio of the interaction volume to the cell volume:
\[
    P_{\text{base}} = \frac{V_{\text{int}}}{V_{\text{cell}}}.
\]

Two objects whose AABBs overlap need not intersect: if most of each
bounding box is empty (e.g., elongated vessels or thin sheet-like structures),
the actual mesh-level intersection probability is far below what the AABB
volume ratio suggests~\cite{ltree}. We correct this with a per-cell shape factor derived
from the volume ratios. Combining both datasets symmetrically through the geometric mean,
\[
    R_{\text{combined}} = \sqrt{\mathrm{VolRatio}_1 \cdot \mathrm{VolRatio}_2},
\]
we apply a power law with a tunable exponent $\gamma$
(default $0.8$) that controls how strongly sparse structures are down-weighted:
\[
    F_{\text{shape}} = R_{\text{combined}}^{\gamma}.
\]

For large objects in a fine grid, $V_{\text{int}}$ can exceed $V_{\text{cell}}$ and push the product above $1$.
Therefore, we cap it at $1.0$ to obtain the final per-cell intersection probability:
\[
    P = \min\!\left(P_{\text{base}} \cdot F_{\text{shape}},\; 1.0\right).
\]

Approximating every candidate pair in the cell by the per-cell average sizes, the expected number of intersecting pairs contributed by the cell is
\[
  E_i = N_1 \cdot N_2 \cdot P,
\]
and the selectivity estimate sums these per-cell expectations over all occupied cells:
\[
  E_{\text{final}} = \sum_i E_i.
\]

The final estimate $E_{\text{final}}$ is used to size the query's deduplication hashmap. In the current implementation, the target number of slots is obtained by dividing $E_{\text{final}}$ by a hash-load factor of $0.5$. This keeps the table sparse even when the estimate runs somewhat low, so collision chains stay short.

\noindent\textbf{Handling containment.} The overlap and intersection joins can use either
strategy, but the containment join cannot use the counting strategy. For each candidate pair, the parity test of Algorithm~\ref{alg:containment} must check whether the pair
has already been recorded as a surface crossing, which requires $H_{cross}$ to be a queryable
hashmap rather than a flat buffer. Therefore, the containment join allocates two
hashmaps, $H_{cross}$ for surface crossings and $H_{contain}$ for containment results, both sized by the estimator above.

\subsection{Discussion}
\label{sec:discussion}

\noindent \textbf{Robustness.}
\sysname assumes closed, watertight manifold inputs where each interior edge is shared by exactly two faces.
Non-manifold inputs, such as open surfaces or T-junctions, fall outside the standard spatial join semantics for closed polyhedra and are typically repaired upstream by mesh-cleaning tools.
\sysname reports a crossing only when an edge pierces a face, so purely tangential contact, where two surfaces touch but neither interior enters the other, is outside the overlap and intersection semantics.
During tracing, FP32 rounding can make it ambiguous whether a ray grazing a shared edge hits one, both, or neither of the adjacent triangles.
OptiX's built-in triangle intersector resolves this by reporting a ray crossing a surface at or near a shared edge exactly once~\cite{woop2013}, even in single precision, so it neither leaks through a crack between adjacent faces nor records the same crossing twice, which is what the parity test requires.
Within the multi-hit loop, \sysname advances the lower ray bound with \textsc{NextAfter} to the next representable FP32 value, ensuring single crossings are not re-reported on the following iteration.
Degenerate configurations, such as zero-area triangles, can produce unreliable FP32 results and require preprocessing or exact arithmetic.

\noindent \textbf{Out-of-GPU-core execution.}
\sysname assumes that the BVHs of both datasets and the result structure fit within the GPU's memory.
Workloads that exceed the device's capacity could be handled using a tiled execution scheme.
A coarse spatial grid partitions the shared space of both datasets, and each object is replicated to every cell it intersects, after which \sysname joins the two datasets cell by cell.
The adaptation is straightforward, as each cell reuses the same kernels and BVH layout.
Only the grid partitioning and cell iteration need to be added, while the core ray-tracing primitives remain the same.
We leave this implementation to future work, as our evaluation targets workloads that fit in device memory.

\noindent \textbf{Updates.}
The workloads targeted by \sysname, such as digital pathology and connectomics, are predominantly read-heavy: meshes are reconstructed once and queried many times.
For dynamic scenes, OptiX can update an acceleration structure in place if the structure was built with update support and the number of primitives remains constant.
This can work well for small deformations or repositioning of existing geometries.
Larger deformations or topology changes are better handled by rebuilding the affected BVHs, as repeated in-place updates gradually degrade traversal quality.
We leave extending \sysname to dynamic workloads as future work.

\section{Experimental Evaluation}
\label{sec:experimental-evaluation}

We evaluate \sysname against state-of-the-art baselines, study its scalability with dataset size, mesh complexity, and result selectivity, and break down the execution costs for its three join predicates.

\subsection{OptiX Implementation}
\label{sec:exp:optix}

\sysname is implemented in C++ and CUDA with NVIDIA OptiX 7.5 and CUDA 12.8. Each dataset is represented as OptiX triangle primitives with face culling disabled (\texttt{\seqsplit{OPTIX\_GEOMETRY\_FLAG\_DISABLE\_TRIANGLE\_FACE\_CULLING}}), and its acceleration structure is built as a single GAS with \texttt{OPTIX\_BUILD\_FLAG\_NONE}. For geometries traversed by any-hit programs, the triangle input additionally uses \texttt{\seqsplit{OPTIX\_GEOMETRY\_FLAG\_REQUIRE\_SINGLE\_ANYHIT\_CALL}} to guarantee at most one any-hit invocation per primitive intersection.
All OptiX pipelines are configured for a single-GAS traversable graph (\texttt{\seqsplit{OPTIX\_TRAVERSABLE\_GRAPH\_FLAG\_ALLOW\_SINGLE\_GAS}}). The overlap and edge-based phases use raygen/closest-hit program groups, while the containment parity path additionally uses an any-hit program. Output hashmaps are sized from precomputed uniform-grid statistics evaluated by a lightweight CUDA kernel.

\subsection{Experimental Setup and Methodology}
\label{sec:exp:setup}

\textbf{Artifact Availability.} To support reproducibility, the \textsc{Pierce} source code, evaluation datasets, and evaluation framework are publicly available.\footnote{Source code:
\url{https://github.com/AntonHackl/Pierce}}
\footnote{Datasets: \url{https://huggingface.co/HPI-Pierce}}

\noindent\textbf{Hardware Configuration.} All experiments were performed on a machine equipped with 20 Intel Xeon 6517 CPU cores, 100~GB RAM, and an NVIDIA RTX PRO 6000 GPU (96~GB VRAM, 188 RT cores). All C++ components were compiled in Release mode with GCC 13.3 and \texttt{-O3}. All datasets reside on an NFS volume.

\noindent\textbf{Datasets.}
For our experiments, we use three dataset families, summarized in Table~\ref{tab:dataset_benchmark}.
\textsc{Tissue} consists of synthetic brain-tissue meshes produced by the data simulator of TDBase~\cite{tdbase, tdbase-simulator}, which models the geometry of structures recovered from serial-section electron microscopy.
This allows us to evaluate \sysname on the same data that TDBase was designed for.
We evaluate two types of joins. The first joins nuclei with vessels, identifying the nuclei located within tissue irrigated by vessels. 
The second joins two independently generated nuclei populations.
\textsc{MICrONS} contains neuron meshes from the cortical millimeter connectome~\cite{microns}, a reconstruction of mouse visual cortex. Neurons are heavily branched, tree-like structures with extensive dendritic and axonal arbors, producing larger and more irregular meshes than \textsc{Tissue}'s. 
We create four neuron datasets that we join pairwise by first extracting two dense spatial subsets from the full MICrONS reconstruction and then partitioning the neurons in each subset into two disjoint populations of roughly equal size via an alternating split.
Finally, we use \textsc{Synthetic} cube and tessellated-sphere data to vary individual workload parameters, such as mesh complexity and selectivity.

\begin{table*}[t]
\centering
\scriptsize
\setlength{\tabcolsep}{4pt}
\begin{tabular}{lp{5.1cm}rrrrr}
\toprule
Dataset & Description & \#Objects & \#Triangles & Size & Preproc.\ (s) [Pierce/TDBase] & Loading (ms) [Pierce/TDBase] \\
\midrule
\textit{\textsc{Tissue}} & Brain-tissue meshes from the TDBase simulator (\url{https://github.com/tengdj/tdbase}) & & & & & \\
\quad Nuclei$_1$ & Cell nuclei   & 583{,}200 & 172.6M & 3.90 GB & 512.20 / 64469.54 & 70.17 / 6939.57 \\
\quad Vessel & Blood vessels  & 729 & 23.4M & 495.02 MB & 78.14 / 21880.06 & 11.56 / 290.04 \\
\quad Nuclei$_2$ & Cell nuclei  & 291{,}600 & 86.3M & 1.95 GB & 234.90 / 32343.20 & 35.30 / 2428.86 \\
\quad Nuclei$_3$ & Cell nuclei, independent sample, same parameters as Nuclei$_2$ & 291{,}600 & 86.3M & 1.95 GB & 231.67 / 32241.85 & 35.23 / 2435.64 \\
\midrule
\textit{\textsc{MICrONS}} & Neuron meshes, regional subsets of the cortical connectome (\url{https://www.microns-explorer.org/}) & & & & & \\
\quad Neurons$_1$ & Neuron meshes & 20 & 127.6M & 6.26 GB & 207.74 / -- & 66.11 / -- \\
\quad Neurons$_2$ & Neuron meshes & 19 & 115.6M & 5.67 GB & 185.45 / -- & 60.24 / -- \\
\quad Neurons$_3$ & Neuron meshes & 38 & 233.6M & 11.62 GB & 406.65 / -- & 94.89 / -- \\
\quad Neurons$_4$ & Neuron meshes & 37 & 242.0M & 12.06 GB & 428.60 / -- & 98.17 / -- \\
\midrule
\textit{\textsc{Synthetic}} & Generated meshes for controlled parameter sweeps & & & & & \\
\quad Cubes$_1$ & Uniformly sampled axis-aligned cubes & 200{,}000 & 2.4M & 110.95 MB & 4.08 / -- & 1.42 / -- \\
\quad Cubes$_2$ & Uniformly sampled axis-aligned cubes & 1{,}000{,}000 & 12.0M & 570.19 MB & 21.17 / -- & 5.57 / -- \\
\quad Spheres$_1$\textsuperscript{\dag} & Tessellated spheres & 500 & 41.2M & 1.78 GB & 75.45 / 47.05 & 16.98 / 12665.38 \\
\quad Spheres$_2$\textsuperscript{\dag} & Tessellated spheres & 500 & 41.2M & 1.78 GB & 78.12 / 47.57 & 16.95 / 12970.25 \\
\bottomrule
\end{tabular}
\vspace{2pt}
\parbox{\linewidth}{\footnotesize{$^\dagger$ For spheres, TDBase uses RAW-format \texttt{.dt} files, since its simulator outputs compressed files only for the \textsc{Tissue} datasets. We produced these files with a custom extension of the TDBase preprocessing pipeline. As the data is uncompressed, the timings reflect a compatible but not optimal TDBase preprocessing and loading pipeline.}}
\caption{Datasets used in our evaluation, with preprocessing and loading times. A dash indicates that TDBase does not support the dataset.}
\label{tab:dataset_benchmark}
\end{table*}

\noindent \textbf{Baselines.} We compare \sysname (in its default configuration, which performs selectivity estimation) against three baselines for the overlap join:
(i) \textbf{TDBase}~\cite{tdbase}, the state-of-the-art GPU-accelerated 3D spatial join technique, which follows the filter-and-progressive-refine paradigm. TDBase compresses each polyhedron via progressive protruding-vertex pruning into a sequence of levels of detail (LODs). At query time, refinement starts from the coarsest LOD and proceeds to higher resolutions only when the lower-resolution geometry is insufficient to decide the predicate. The filter stage runs on the CPU with OpenMP-based multi-threading (20 threads), while face-pair geometric evaluation in the refinement stage is offloaded to the GPU via CUDA. 
We use the authors' original implementation\footnote{\url{https://github.com/tengdj/tdbase}} with all LODs enabled.
Both TDBase and its mesh simulator operate in single-precision (FP32), as does \sysname{}.
(ii) \textbf{TOUCH}~\cite{nobari2013touch}, an in-memory spatial join that uses hierarchical data-oriented partitioning. 
It builds an R-tree for one of the input datasets and assigns each object from the second dataset to a single node of that tree. Each object from the second dataset is then compared only with objects from the first dataset that are indexed within that node's subtree. This approach helps to keep the candidate pair count low, even as the data density increases.
TOUCH is implemented in C++.
(iii) \textbf{Face}, a parallel CPU baseline we implemented based on BVH indexing. For overlap, it builds a BVH over the triangles of the first input and queries it with triangles from the second to find overlapping object pairs.
For intersection and containment, it additionally performs point-in-mesh tests. 
Face is implemented in C++ with CGAL 5.6.1 and OpenMP multi-threading (20 threads). 
We also use Face as a reference to cross-check \sysname's output: across all evaluated workloads and all three predicates, the two report the same number of result pairs.

\noindent \textbf{Preprocessing.} 
For \sysname, preprocessing converts each input dataset into a flat triangle representation with per-triangle object mappings, extracts the corresponding per-object edge set used during ray generation, and serializes the result to disk for reuse. 
It also partitions the dataset space into a uniform grid and materializes sparse per-cell statistics for selectivity estimation, including object occupancy, average object size, and an approximate volume ratio, i.e., the fraction of an object's bounding-box volume occupied by the mesh.
For TDBase, preprocessing applies progressive protruding-vertex pruning to each polyhedron, producing a compressed multi-LOD representation. 
As this progressive compression is costly, \sysname preprocesses the \textsc{Tissue} datasets $\sim$126--280$\times$ faster than TDBase. 
TOUCH and Face have no offline preprocessing step. They build their in-memory indexes at query setup.

\noindent \textbf{Loading.} 
We define loading as the cost of bringing the preprocessed data into a query-ready in-memory state.
\sysname uploads the triangle data, per-triangle object mappings, and precomputed edge sets to GPU memory and builds the per-dataset OptiX acceleration structures.
TDBase decodes the compressed multi-LOD representation produced by preprocessing into its in-memory working format.
As this mesh decoding is costly, \sysname loads the \textsc{Tissue} datasets $\sim$25--99$\times$ faster than TDBase.
TOUCH and Face persist no data structures, so they have no loading cost.

We report preprocessing and loading times separately from query time, since they are one-time setup costs amortized over multiple queries.
Table~\ref{tab:dataset_benchmark} lists these times for each dataset, where dashes mark datasets that TDBase does not support.
Specifically, TDBase does not support our \textsc{Synthetic} cube datasets because it cannot generate LODs for them. The required object-compression step during preprocessing fails.
Likewise, it does not support our \textsc{MICrONS} datasets because, during preprocessing, it invokes CGAL routines whose input constraints are not satisfied by these neuron meshes.
For the synthetic sphere datasets, TDBase input tiles were generated using a custom extension of the original TDBase preprocessing code. In particular, we produced TDBase RAW-format \texttt{.dt} files compatible with the sphere benchmark instances used in our evaluation. Consequently, the reported TDBase preprocessing and loading times for the sphere datasets reflect this adapted compatibility-oriented pipeline and should not be interpreted as timings from an optimized native TDBase preprocessing workflow.

\noindent \textbf{Measurement.} 
Across all approaches, the reported query time measures join execution only, excluding preprocessing, loading, and index or acceleration-structure construction.
Each reported query time is the mean of five runs after three warm-up runs.

\begin{figure*}[t]
  \centering
  \begin{minipage}[t]{0.32\linewidth}
    \centering
    \includegraphics[width=\linewidth]{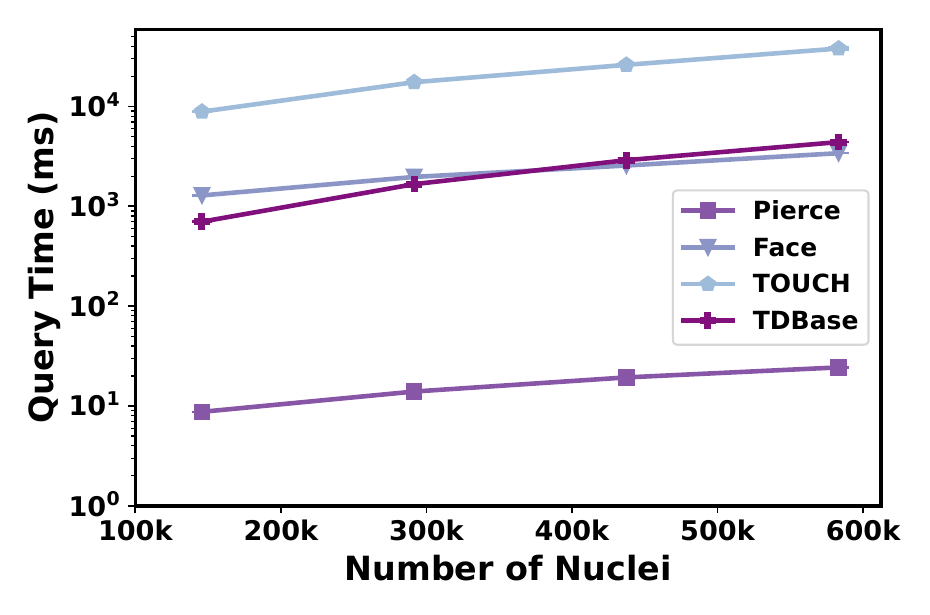}
    \caption*{(a) Scaling with input size}
  \end{minipage}
  \hfill
  \begin{minipage}[t]{0.32\linewidth}
    \centering
    \includegraphics[width=\linewidth]{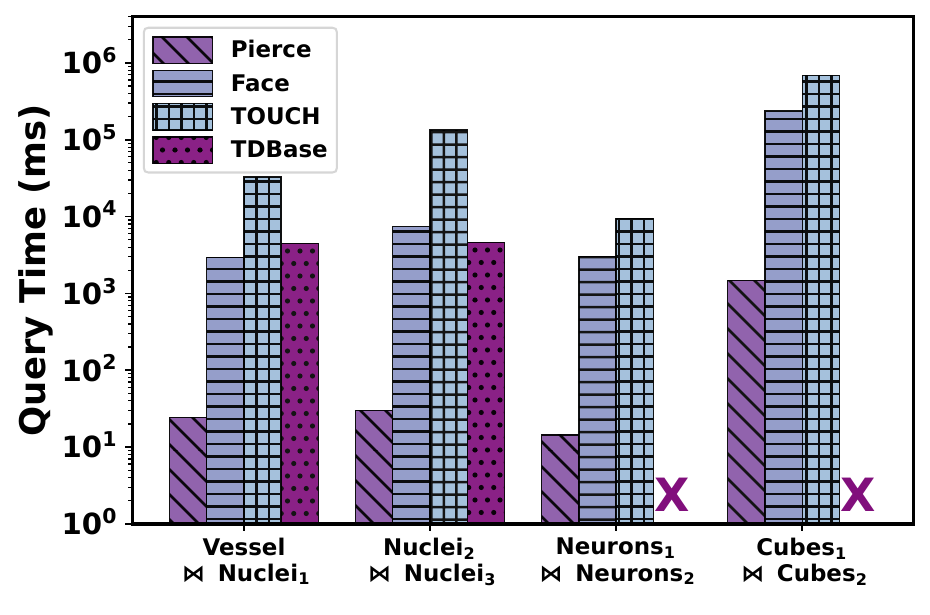}
    \caption*{(b) Comparison across workloads}
  \end{minipage}
  \hfill
  \begin{minipage}[t]{0.32\linewidth}
    \centering
    \includegraphics[width=\linewidth]{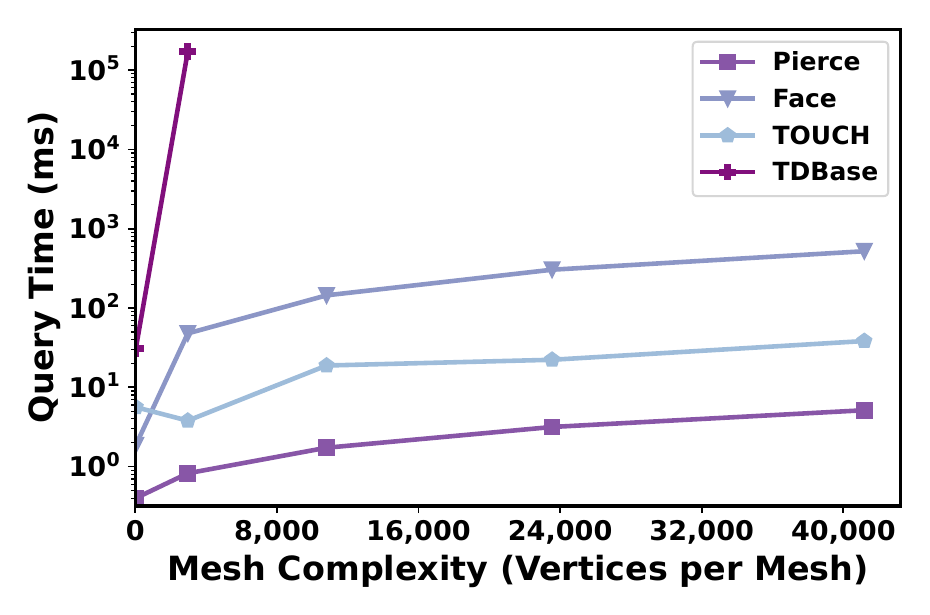}
    \caption*{(c) Scaling with mesh complexity}
  \end{minipage}
  \caption{Overall performance evaluation for the overlap join: (a) scalability with increasing input size on \textsc{Tissue} data, (b) performance across different workloads, and (c) scalability with increasing mesh complexity.}
  \Description{Three logarithmic runtime plots show Pierce consistently outperforming Face, TOUCH, and TDBase by large margins. In panel (a), as the number of nuclei grows from about 146,000 to 583,000, Pierce rises only from roughly 8 to 24 milliseconds, while all baselines require seconds or tens of seconds at the largest size. In panel (b), Pierce has the lowest runtime across four workloads; TDBase is unavailable for the neuron and cube workloads. In panel (c), increasing mesh complexity to about 41,000 vertices per mesh raises Pierce only to about 5 milliseconds, compared with about 38 milliseconds for TOUCH and about 520 milliseconds for Face; TDBase exceeds the plotted range after the second measurement.}
  \label{fig:overall-performance}
\end{figure*}

\subsection{Overall Performance}
\label{sec:exp:overall}
Figure~\ref{fig:overall-performance} reports \sysname's query time for the overlap join under varying input size, workload, and mesh complexity.

Figure~\ref{fig:overall-performance}(a) shows execution time for the \texttt{Nuclei$_1$}~$\bowtie$~\texttt{Vessel} join as we scale the nuclei dataset from $145,800$ to $583,200$ nuclei while keeping the vessel dataset fixed at $729$ vessels. 
\sysname is the only approach that remains interactive across the entire evaluated range, completing the largest join in ${\sim}24$~ms, while the baselines require ${\sim}3.4$~s (Face), ${\sim}4.4$~s (TDBase), and ${\sim}38.1$~s (TOUCH).
Moreover, \sysname's advantage increases with input size: quadrupling the nucleus count increases its runtime by ${\sim}2.8\times$, versus ${\sim}6.3\times$ for TDBase, ${\sim}4.3\times$ for TOUCH, and ${\sim}2.6\times$ for Face. Despite Face's somewhat lower growth, it remains far from interactive, with average query time remaining above one second throughout.
\sysname's speedup over the state-of-the-art GPU-accelerated TDBase grows to ${\sim}181\times$ at the largest input.

Figure~\ref{fig:overall-performance}(b) reports query time for the overlap join across four representative workloads spanning the three dataset families.
These workloads effectively test the two main factors that affect 3D spatial joins: mesh complexity and object count.
\textsc{MICrONS} isolates the first factor with few but complex meshes, \textsc{Cubes} isolates the second at the million-object scale with simple per-object geometry, and \textsc{Tissue} combines both, featuring moderate complexity with large object counts.
\sysname achieves the lowest query time across all workloads.
Compared to the next-fastest baseline for each workload, it is ${\sim}140\times$ faster on \texttt{Vessel} $\bowtie$ \texttt{Nuclei$_1$} (Face), ${\sim}150\times$
on \texttt{Nuclei$_2$}~$\bowtie$~\texttt{Nuclei$_3$} (TDBase), ${\sim}160\times$ on \texttt{Neurons$_1$} $\bowtie$ \texttt{Neurons$_2$} (Face), and ${\sim}180\times$ on \texttt{Cubes$_1$} $\bowtie$ \texttt{Cubes$_2$} (Face).
This consistently demonstrates speedups of more than two orders of magnitude.
The performance margin is even larger against the slower baselines, reaching ${\sim}3{,}500\times$ over TOUCH on \texttt{Nuclei$_2$}~$\bowtie$~\texttt{Nuclei$_3$}.
TDBase does not support the \textsc{MICrONS} or \textsc{Cubes} workloads (see Section~\ref{sec:exp:setup}) and is therefore absent from the corresponding bars.

\subsection{Scalability with Mesh Complexity}
\label{sec:exp:mesh_complexity}

This experiment evaluates how the approaches scale with mesh complexity while keeping all other factors fixed.
We use the \textsc{Synthetic} sphere datasets, fixing the placement of 500 spheres per dataset and varying the tessellation of the shared sphere template across five stages, from 26 to 41{,}186 vertices (48 to 82{,}368 triangles) per sphere.
At the dataset level, this grows each relation from 24{,}000 to 41.18\,M triangles.
Since the placement remains unchanged, the result size is constant at 100 pairs.

Figure~\ref{fig:overall-performance}(c) reports the results.
\sysname achieves the lowest query time across the complexity range by mapping both filtering and refinement onto hardware-accelerated ray tracing. The number of edge rays it casts grows linearly with mesh complexity, while the cost of each ray grows only logarithmically with the size of the target BVH. As a result, \sysname scales near linearly with mesh complexity, achieving low query latency even for complex meshes. 

TDBase exhibits the steepest growth: it increases from 30.6\,ms at 26 vertices to 171.6\,s already at 2{,}966 vertices, and then exceeds our $5$-minute timeout at higher mesh complexities (from 10{,}806 vertices onward).
In this sphere workload, TDBase's object- and voxel-level candidate counts remain nearly constant, but the number of triangle pairs inside those candidate voxel pairs grows roughly quadratically with tessellation.
As a result, geometric computation quickly dominates TDBase's query time even though TDBase, like \sysname, offloads this computation to the GPU.
However, TDBase evaluates triangle pairs on general-purpose CUDA cores, whereas \sysname maps the equivalent computation to the dedicated RT cores built for ray–triangle intersection.

Similar to \sysname, Face finds intersecting objects by traversing a BVH.
However, Face queries the BVH with triangles and evaluates triangle--triangle tests at the leaves on the CPU, whereas \sysname queries with rays and evaluates ray--triangle tests.
Since the GPU's RT cores are built specifically for the ray--triangle primitive, \sysname performs this search in dedicated hardware, while Face runs the equivalent triangle--triangle search on general-purpose CPU cores, which makes Face about $101\times$ slower than \sysname at 41{,}186 vertices.

Unlike Face, TOUCH first filters at the object level: its hierarchical, data-oriented partitioning restricts refinement to the object pairs whose subtrees overlap, rather than sweeping every triangle of one dataset through the other's hierarchy.
With only 500 objects per dataset, few pairs survive this filter, so TOUCH does little refinement and is faster than Face, increasing from 5.62\,ms to 38.35\,ms.
However, since refinement still performs triangle--triangle tests on the CPU for each candidate pair rather than ray tracing on RT cores, TOUCH is about $7.5\times$ slower than \sysname.

\subsection{Join Predicate Comparison and Breakdown}
\label{sec:exp:other-predicates}

\begin{figure}[t]
  \centering
  \includegraphics[width=\linewidth]{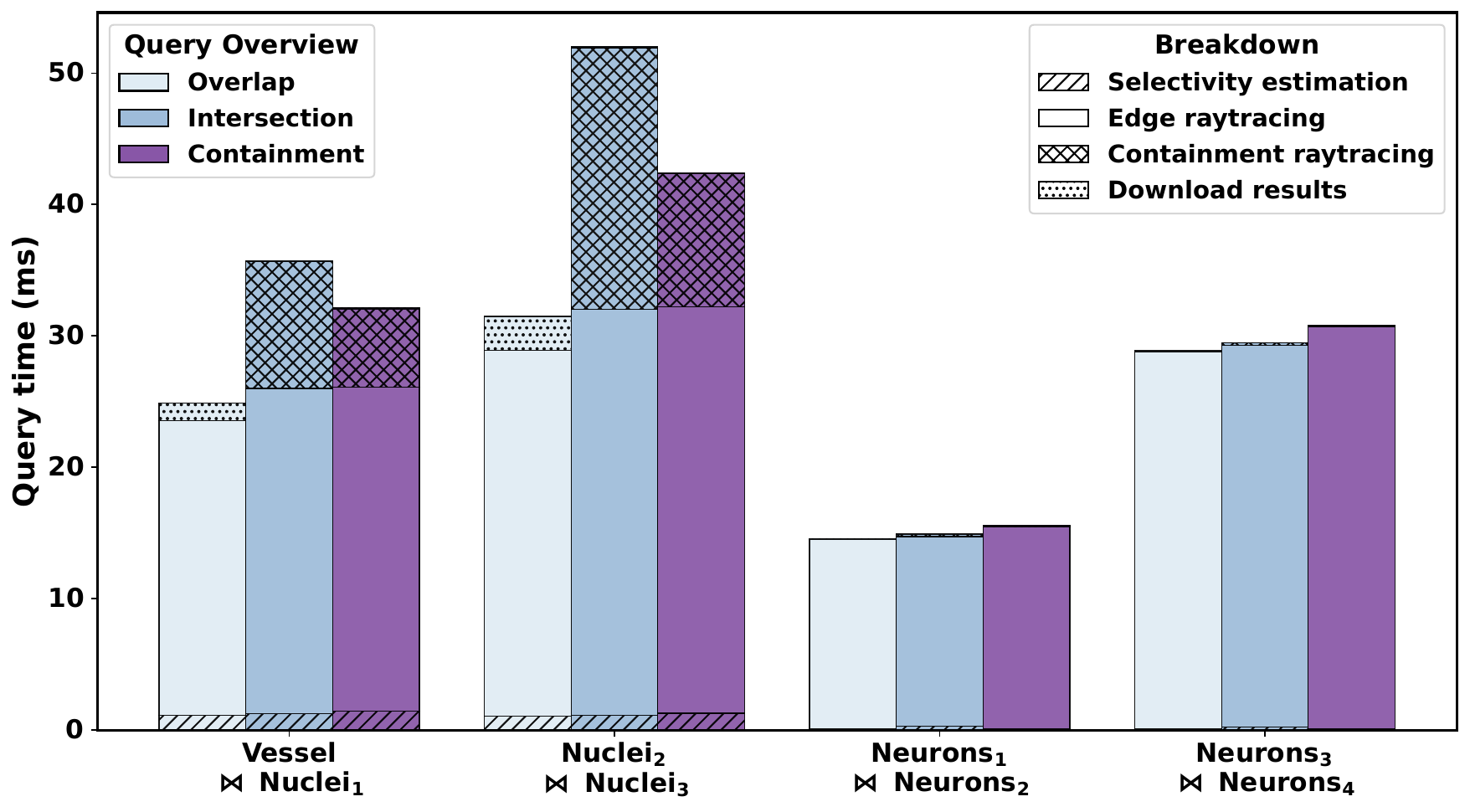}
  \caption{Comparison of query runtimes across different workloads and join predicates.}
  \Description{Grouped stacked bars compare overlap, intersection, and containment runtimes on four workloads. Overlap is lowest for every workload, ranging from about 14 to 31 milliseconds, because it consists almost entirely of edge ray tracing. On the two Tissue workloads, the additional containment-ray-tracing phase increases intersection and containment runtimes to roughly 32--52 milliseconds; the largest increase occurs for intersection on the dense Nuclei 2, Nuclei 3 join workload. On the two MICrONS workloads, all three predicates remain close together at about 14--31 milliseconds because containment tracing adds little cost. Selectivity estimation and result download contribute only thin portions of each bar.}
  \label{fig:predicate_breakdown}
\end{figure}

This section analyzes the cost of \sysname's three spatial join predicates, i.e., overlap, intersection, and containment
(Sections~\ref{sec:overlap}--\ref{sec:intersection}), and where
execution time is spent.\footnote{We restrict this analysis to \sysname. The TDBase paper describes an intersection variant, but the publicly available code implements only overlap.}
Figure~\ref{fig:predicate_breakdown} reports query time for the three predicates on two \textsc{Tissue} and two \textsc{MICrONS} joins.
We break down the query time into four phases: selectivity estimation, bidirectional edge ray tracing, containment ray tracing (the point-in-mesh parity test of Algorithm~\ref{alg:containment}), and the final transfer of the result set from GPU to host memory (labeled as download results).

Among the three predicates, overlap is the only one that does not perform containment ray tracing, and thus incurs the lowest query time across all workloads, from $\sim$14.5~ms on $\texttt{Neurons}_1 \bowtie \texttt{Neurons}_2$
to $\sim$31.5~ms on $\texttt{Nuclei}_2 \bowtie \texttt{Nuclei}_3$.
Its cost is dominated by bidirectional edge ray tracing, with selectivity estimation and CPU--GPU result transfer adding minimal overhead.

Intersection and containment additionally perform point-in-mesh parity tests, whose cost scales with the number of rays cast (determined by the cardinality of $D_2$) and the number of hits per ray (which increases with $D_1$'s density).
In the two \textsc{MICrONS} workloads ($|D_2| = 19$ and $|D_2| = 37$), so few rays are cast that the parity test adds negligible overhead.
On $\text{Vessel} \bowtie \text{Nuclei}_1$, many more rays are cast, but the vessel dataset is sparse, so each ray hits few objects, and the parity test adds only $\sim$7.2--10.8~ms over overlap.
In contrast, on $\text{Nuclei}_2 \bowtie \text{Nuclei}_3$, the rays are cast into a dense nuclei BVH, and the parity test adds $\sim$20.5~ms to intersection and $\sim$10.9~ms to containment over the overlap baseline.
The gap between the two predicates arises from how they handle the parity results.
Containment checks each odd-parity pair against $H_{\text{cross}}$ (Algorithm~\ref{alg:containment}, line~10), discards pairs whose surfaces cross, and writes the survivors into a separate containment hashmap. 
Intersection applies no such filter: it records \emph{every} odd-parity pair into a shared result hashmap that already holds all surface-crossing pairs. 
On dense \textsc{Tissue} data, where many pairs satisfy both conditions, intersection performs far more insertions into a fuller table, making its containment ray tracing phase more expensive.
This ordering reverses on \textsc{MICrONS}: with only a handful of result pairs, intersection has almost nothing extra to record, so the performance advantage vanishes and containment is marginally slower (by $\sim$0.5--1.5~ms) owing to its additional $H_{\text{cross}}$ lookup per candidate.

Selectivity estimation contributes a small fixed cost across all workloads ($\lesssim$2.5~ms).
Result download grows with the size of the result set but remains under $9\%$ of total query time, even for the largest result set.

\begin{figure}[t]
  \centering
  \includegraphics[width=\linewidth]{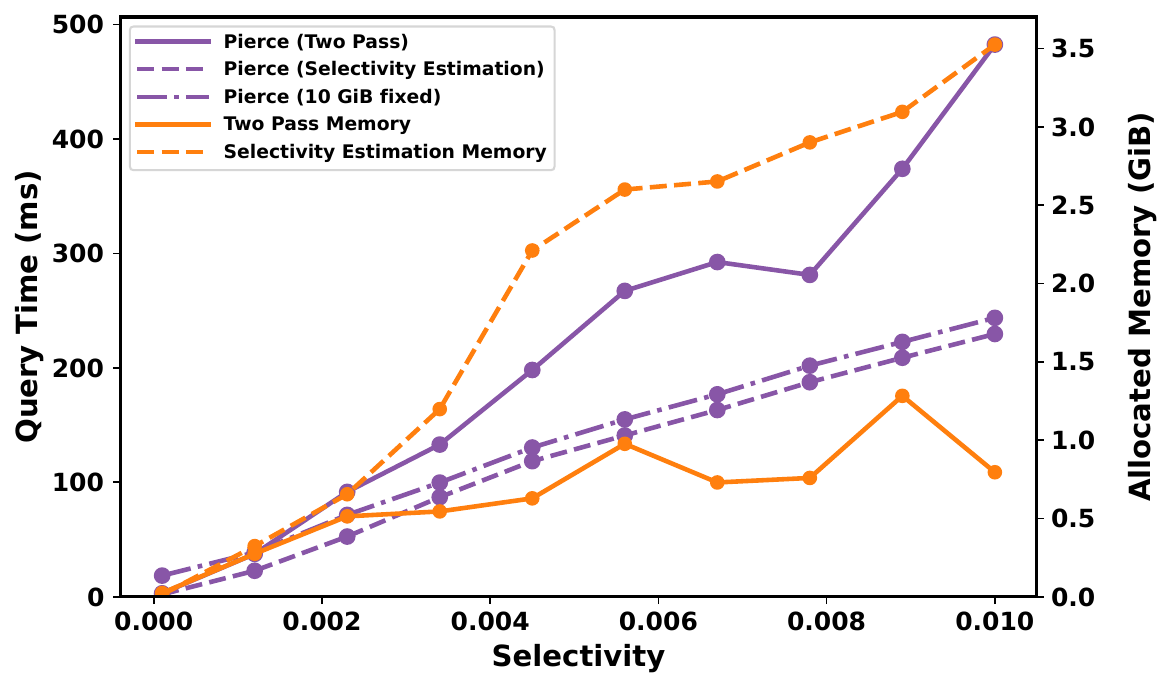}
   \caption{Query time (left) and allocated output memory (right) of three \sysname variants as join selectivity increases.}  
   \Description{Dual-axis line chart comparing three Pierce materialization variants over selectivity from 0.0001 to 0.01. The selectivity-estimation variant has the lowest query time throughout, reaching about 225 milliseconds at the highest selectivity, while the fixed 10 GiB variant is slightly slower and the two-pass variant grows to about 480 milliseconds. Allocated memory for selectivity estimation increases steadily with selectivity to about 3.5 GiB, remaining well below the fixed variant's constant 10 GiB allocation; two-pass memory is generally lower but varies non-monotonically.}
   \label{fig:selectivity-scaling}
\end{figure}

\subsection{Impact of Selectivity Estimation}
\label{sec:exp:estimation}

This section studies the impact of selectivity estimation and scalability with join selectivity, comparing three variants.
The first two are the materialization strategies of Section~\ref{sec:result-collection}: \emph{Two-Pass} (\emph{Two Pass} in Figure~\ref{fig:selectivity-scaling}) counts the number of ray–triangle hits in a first tracing pass before emitting records into a flat buffer, while \emph{Estimated} (our default, labeled \emph{Selectivity Estimation} in Figure~\ref{fig:selectivity-scaling}) inserts pairs into a hashmap sized by the Minkowski-histogram estimator in a single pass.
The third, \emph{Fixed} (\emph{10 GiB fixed} in the figure), skips estimation and sizes the hashmap to a fixed fraction of GPU memory.
For each variant, we sweep the overlap-join selectivity from $10^{-4}$ to $10^{-2}$ on two synthetic cube datasets of 50{,}000 cubes each, reporting query time and allocated output memory.
Selectivity is controlled by varying the extent of the cubic universe while keeping the cube count and size range fixed.

The \emph{Estimated} variant is the fastest across the entire selectivity range. 
\emph{Fixed} is only marginally slower: both perform a single tracing pass and differ only in hashmap capacity. 
In our implementation, this larger table increases several costs: it is more expensive to allocate and initialize, and the final compaction step must scan more slots even when many remain empty. \emph{Estimated} avoids this overhead by sizing the hashmap closer to the actual result cardinality.
The \emph{Two-Pass} variant traverses every edge ray twice, once to count and once to emit, roughly doubling the traversal cost of the single-pass variants.
This overhead grows with selectivity, as both the extra hit processing and the sort-based deduplication of the flat buffer scale with the number of results.
It is therefore the slowest variant except at very low selectivity.
There, few results are produced, so the extra counting pass and flat-buffer deduplication remain cheap, whereas \emph{Fixed} still pays the overhead of operating on an oversized hash table.

The right axis shows the allocated output memory for the \emph{Two-Pass} and \emph{Estimated} variants.
\emph{Fixed} allocates a constant $10$~GiB hashmap by design. 
In comparison, \emph{Estimated}'s allocation peaks at $3.5$~GiB at the highest selectivity, roughly three times smaller than \emph{Fixed}, freeing GPU memory for larger workloads. 
The $0.5$ load factor (Section~\ref{sec:result-collection}) sizes \emph{Estimated}'s hashmap to twice the estimated number of result pairs.
At the highest selectivity, \emph{Estimated} uses about $4.4\times$ the $\sim 0.8$~GiB allocated by \emph{Two-Pass}.
We deliberately favor a larger hashmap: overestimating its required size only incurs additional memory overhead, whereas underestimating it lengthens collision chains and can force a kernel restart. 
The two memory curves nearly coincide at low selectivity and separate as selectivity increases.
This is because \emph{Estimated} allocates both a hash table and an output buffer that grow with the estimated result cardinality, whereas \emph{Two-Pass} allocates only the flat result buffer sized by its first counting pass. As selectivity increases, the additional hash-table footprint of \emph{Estimated} becomes increasingly pronounced.
Note that the \emph{Two-Pass} memory allocation does not grow monotonically with selectivity.
This is because the buffer tracks the raw ray–triangle hit count, not the number of result pairs that selectivity measures, and a single pair can produce several hits that are deduplicated only at the end.

Overall, \sysname's selectivity estimator achieves the best of both worlds.
It has the lowest query time of the three variants and a compact memory footprint, whereas \emph{Two-Pass} sacrifices the former and \emph{Fixed} the latter. 
\emph{Estimated} is fastest because it avoids both the extra tracing pass of \emph{Two-Pass} and the oversized hashmap overheads of \emph{Fixed}.

\section{Related Work}
We group related work into three categories: 3D polyhedral spatial join techniques, repurposing GPU graphics hardware for spatial workloads, and using ray-tracing cores for database workloads.

\noindent \textbf{3D Spatial Join over Polyhedral Objects.}
Although 2D spatial joins have a rich literature spanning hash-based~\cite{lo1996spatial}, tile-based~\cite{patel1996partition}, approximation-based~\cite{rasterapprox, georgiadis2025raster}, and distributed methods~\cite{yu2015geospark, baig2017sparkgis}, the 3D polyhedral case has received far less attention.
Existing techniques follow the filter-and-refine paradigm and accelerate (one of) its two stages: the filtering step that prunes candidate object pairs and the refinement step that evaluates the exact predicate over mesh triangles.
The first direction strengthens the filter step.
For example, TOUCH~\cite{nobari2013touch} accelerates filtering through hierarchical data-oriented partitioning of object bounding boxes and naturally extends to 3D, but does not address mesh-level refinement.
The second direction targets refinement, which dominates query time on complex meshes.
3DPro~\cite{3dpro} introduces the filter-and-progressive-refine paradigm with protruding-vertex pruning compression and parallelizes face-pair evaluation on CUDA cores. TDBase~\cite{tdbase} extends 3DPro with facet-level Hausdorff and proxy Hausdorff bounds, yielding tighter bounds at low LODs, making it the state-of-the-art GPU-accelerated approach for 3D polyhedral spatial joins.
Concurrent with our work, 3DPipe~\cite{yuan2026threedpipe} parallelizes both the voxel-pair filtering and facet-level refinement stages of the filter-and-progressive-refine pipeline on the GPU.
\sysname{} departs from the above approaches by replacing the filter-and-refine pipeline with a hardware-accelerated BVH traversal that both prunes distant geometry (filter) and detects surface crossings (refine).
The deeper distinction lies in the primitive. 
Prior work refines each candidate pair through pairwise triangle--triangle tests on general-purpose cores, whereas \sysname{} replaces these with ray--triangle tests on dedicated RT cores.

\noindent \textbf{Spatial Query Processing on GPUs.}
Over the past decade, several research efforts have leveraged programmable GPUs to accelerate spatial queries.
A first line of work~\cite{raster-join, spade, gpu-algebra} repurposes the GPU rasterization pipeline for spatial query processing.
For example, RasterJoin~\cite{raster-join} computes 2D spatial aggregations between points and polygons. Furthermore, Doraiswamy and Freire~\cite{spade, gpu-algebra} propose a spatial data model and an algebra designed to exploit modern GPUs.
A more recent line of work~\cite{rtpip, geng2024rayjoin, geng2025librts, wald2019tetmesh} targets RT cores instead.
Wald et al.~\cite{wald2019tetmesh} repurpose RT cores for point location queries in tetrahedral meshes.
Laass~\cite{rtpip} extrudes 2D polygons into 3D walls so that point-in-polygon tests reduce to ray--wall intersections.
RayJoin~\cite{geng2024rayjoin} frames point-in-polygon and line-segment-intersection queries as ray-tracing problems.
LibRTS~\cite{geng2025librts} is a reusable indexing library that packages RT-core spatial query processing, providing a unified BVH traversal for point and range (contains, intersects) queries over 2D data and relieving application programmers from writing workload-specific OptiX code.
In contrast, \sysname targets 3D polyhedral surfaces rather than 2D points and segments.
This necessitates a different ray-construction strategy (edge rays traced bidirectionally) and a different geometric condition (edge--face crossings).

\noindent \textbf{RT Cores for Database Workloads.}
Beyond rendering, RT cores have recently been applied to relational data processing. 
RTIndeX~\cite{henneberg2023rtindex} was the first to use them for column indexing, modeling each value as a triangle along an axis and converting search predicates into rays whose triangle intersections yield the matching rows. 
RTScan~\cite{lv2024rtscan} improved on this design, mapping conjunctive predicate evaluation to ray--primitive intersections in a synthetic 3D space and introducing techniques such as uniform encoding and data sieving to balance ray load. 
This methodology has since evolved from indexing to full query execution~\cite{raydb}.
In contrast, \sysname operates directly on 3D triangle meshes and evaluates true geometric relationships between polyhedra, rather than scalar comparisons over values mapped into a synthetic space. 
To the best of our knowledge, \sysname is the first approach to apply hardware-accelerated ray tracing to spatial joins over 3D polyhedral meshes.

\section{Conclusion}
\label{sec:conclusion}

We present \sysname, an RT-core-accelerated framework for 3D spatial joins over
complex polyhedral meshes.
\sysname reformulates join evaluation as ray tracing, enabling both filtering
and refinement to run on the hardware ray-tracing units of modern GPUs.
Instead of testing triangles pairwise, it casts rays along the unique edges
of one dataset against a BVH built over the other and detects surface
crossings through ray--node and ray--triangle tests.
Additionally, it combines bidirectional tracing with multi-hit traversal to
ensure completeness and introduces a selectivity estimator that efficiently
sizes GPU result structures to optimize performance.
Based on these ideas, \sysname supports overlap, containment, and
intersection joins.
Across digital pathology, connectomics, and synthetic workloads, \sysname
achieves the lowest query time among all evaluated approaches, with over two
orders of magnitude speedup over the state-of-the-art on digital pathology
data.
As future work, we plan to extend \sysname to streaming 3D joins over
dynamic meshes.

\bibliographystyle{ACM-Reference-Format}
\bibliography{references}

\end{document}